\documentclass[fleqn,11pt]{article}

\def\blacksquare{\hbox{\vrule width 4pt height 4pt depth 0pt}}

\title{\bf On the  $d$-complexity of strings}
\author{ZOLT\'AN K\'ASA}

\newcounter{remark}
\newcounter{definition}     \newcounter{bean}
\newcounter{theorem}        \newcounter{lemma}
\newcounter{corollary}      \newcounter{proposition}

\providecommand{\U}[1]{\protect\rule{.1in}{.1in}}

\newtheorem{definition}{Definition}

\newtheorem{proposition}{Proposition}

\newenvironment{references}{\bigskip \begin{center}   {\bf
References} \end{center} \medskip \small
\begin{list}{[\arabic{bean}]}{\usecounter{bean}}}{\end{list}}

\newenvironment{ams}{\smallskip \small{\bf Mathematics
Subject Classification.} }{\medskip }

\date{}

\begin{document}

\maketitle

\begin{center}
{\small\it Faculty of Mathematics and Informatics, 
Babe\c{s}-Bolyai University,\\
RO-3400 Cluj,  str. Kog\u{a}lniceanu 1, Romania,\\
E-mail:  kasa@cs.ubbcluj.ro}
\end{center}

\begin{abstract}
This paper deals with the complexity of strings,
which play  an important role  in  biology  (nucleotid  sequences),
information  theory  and computer science [1,2,4].
The $d$-complexity of a  string  is  defined  as  the number of its distinct
$d$-substrings given in Definition 1. The case $d=1$ is studied in detail. 
\end{abstract}


\section{Introduction}

Let $ X$ be an alphabet, and $X^k$ the set of all strings of length
$k$  over $X.$
The $i$ consecutive appearance of a letter $a$ in a string will be
denoted by $a^i$.
If $i = 0$ then  this  means  the  absence  of  the  corresponding
letter.  The definitions are from [2].

\begin{definition}
Let $d$, $k$ and $s$
be positive integers, {\bf p} $ = x_1x_2\cdots x_k \in  X^k.$
A $d$-substring of {\bf p} is defined as
 {\bf q} = $ x_{i_1}x_{i_2}\cdots x_{i_s}$   where\newline
\hspace*{3 cm}$i_1 \ge  1$,\newline
\hspace*{3 cm}$ 1 \le i_{j+1} - i_j \le d,$ \hspace*{1 cm}
    for $\;\; j = 1,2,\cdots , s-1,$ \newline
\hspace*{3 cm}$i_s \le  k.$
\end{definition}

\begin{definition}
The $d$-complexity ${\bf K}_d({\bf p})$  of the string ${\bf p}$ is the
number of all distinct $d$-substrings of ${\bf p}.$
\end{definition}

{\em Example.} Let ${\it X}$ be the English alphabet and ${\bf p}$ = ISIS. In
this string there are two 2-substrings of length 1  (I, S),
four 2-substrings of length 2 (IS, II, SI, SS), four 2-substrings of length 3
(ISI, ISS, IIS, SIS), and a single
one of length 4 (ISIS). Then ${\bf K}_{2}({\bf p}) = 2 + 4 + 4 + 1 = 11.$
\blacksquare

In the case of strings of length $k$, consisting of different symbols,
the $d$-complexity will be denoted by $N(k,d)$. For any
$k \ge 1$ and ${\bf p} \in  X^k$ we have
$\; k \le  {\bf K}_1({\bf p}) \le  \displaystyle\frac{k(k+1)}{2}.\;$
 If $|X| \ge  2,\;\;  k\ge 1, \;\; d \ge 1$ and ${\bf p} \in X^k$
then $ k \le {\bf K}_d({\bf p}) \le  2^{k}-1.$
\noindent If {\bf p} is a string, consisting of different symbols,
and $ d$ a  positive  integer,
then $ a_{i,d}({\bf p})$ will denote the number of $d$-substrings of {\bf p}
 which  terminate  in the position $ i$. If $k \ge  1$ and ${\bf p} \in X^k$
 consists of different symbols, then for $ \;\;i = 1,2, \ldots ,k$

\begin{equation}
a_{i,d}({\bf p}) = 1 + a_{i-1,d}({\bf p}) + a_{i-2,d}({\bf p}) + \ldots
 + a_{i-d,d}({\bf p}),\hspace*{0.5 cm} 
\end{equation}

\section{Computing the value of  N(k,d)}

The $d$-complexity of a string with different symbols can be obtained
  by  the formula
$$ N(k,d) =\sum_{i=1}^{k}a_{i,d}({\bf p}) $$
where {\bf p} is any string of $k$ different symbols. Because of
(1) we can  write  in the case of $ d \ge  2$
$$  a_{i,d} + \frac{1}{d-1} = \left(a_{i-1,d} + \frac{1}{d-1}\right) + \cdots
 + \left(a_{i-d,d} + \frac{1}{d-1} \right). $$
Let be
$$b_{i,d}=a_{i,d}+\frac{1}{d-1},\hspace*{0.5 cm}{\rm and}\hspace*{0.5 cm}
c_{i,d}=(d-1)b_{i,d}$$
then
$$ c_{i,d} = c_{i-1,d} + c_{ i-2,d}+\ldots + c_{i-d,d}$$
and the sequence $c_{i,d}$ is one of Fibonacci-type.
For any $d$ we have $a_{1,d} = 1$  and from this $c_{1,d} = d$  results.
Therefore the numbers $ c_{i,d}$ are defined by the following recurrence
equations:

\medskip
 $ c_{n,d} = c_{n-1,d} + c_{n-2,d} + \ldots + c_{n-d,d} $
   \hfill for $\;\;n > 0,\hspace*{2 cm} \;$

$ c_{n,d}=1$ \hfill for $\;\; n \le 0.\hspace*{2 cm} \; $

\medskip
These numbers can be generated by the following generating function:

\begin{eqnarray}
F_d(z)&=&\sum_{n\ge 0}^{}{c_{n,d}z^n}=
         \frac{1+(d-2)z - z^2 - \cdots - z^d}{1-2z+z^{d+1}}\nonumber \\
     &=& \frac{1+(d-3)z - (d-1)z^2 +z^{d+1} }{(1-z) (1-2z+z^{d+1})}\nonumber
\end{eqnarray}

The $d$-complexity $N(k,d)$ can be expressed with these numbers $ c_{n,d}$
 by  the following formula:
$$  N(k,d) = \frac{1}{d-1}\left(\sum_{i=1}^{k}{c_{i,d}-k}\right), \;\;\;
\hspace*{0.5 cm}\hbox{for} \;\; d > 1 $$
 and
$$ N(k,1) = \frac{k(k+1)}{2} $$
or
$$
 N(k,d) = N(k-1,d) +\frac{1}{d-1} ( c_{k,d}- 1), \qquad \hbox{for}\;\;\;
 d > 1, \;\; k > 1.$$
If $d = 2$ then
$$ F_2(z) =\frac{1-z^2}{1-2z+z^3} = \frac{1+z}{1-z-z^2}=\frac{F(z)}{z}+F(z) $$
where $F(z)$ is the generating function of the Fibonacci numbers $ F_n$
 (with $F_0= 0,\;\; F_1 = 1$). Then, from this formula we have
$$  c_{n,2} = F_{n+1} +F_{ n} = F_{n+2} $$
and
$$ N(k,2) = \sum_{i=1}^{k}{F_{i+2}}-k = F_{k+4}-k -3 $$
Taking into account  the formula for $F_n$  we have
$$  N (k,2) = \left\lfloor \frac{1}{\sqrt{5}}\left(
\frac{1+\sqrt{5}}{2}\right)^{k+4}+\frac{1}{2}\right\rfloor -k-3$$
which can be approximated by
$$\lfloor 3.0652475\cdot (1.6180339)^k + 0.5\rfloor - k - 3.$$
\medskip
\noindent Table 1 lists the values of $N(k,d)$ for
$k \le  10$ and $d \le  10.$

\begin{center}
\begin{tabular}{|r|rrrrrrrrrr|}\hline
{\small k} $\left\backslash ^d\right.$
          & 1& 2 & 3& 4& 5& 6& 7& 8& 9& 10 \\ \hline
1    & 1& 1& 1& 1& 1& 1& 1& 1& 1& 1 \\
2    & 3& 3& 3& 3& 3& 3& 3& 3& 3& 3 \\
3    & 6& 7& 7& 7& 7& 7& 7& 7& 7& 7 \\
4    & 10& 14& 15& 15& 15& 15& 15& 15& 15& 15 \\
5    & 15& 26& 30& 31& 31& 31& 31& 31& 31& 31 \\
6    & 21& 46& 58& 62& 63& 63& 63& 63& 63& 63 \\
7    & 28& 79& 110& 122& 126& 127& 127& 127& 127& 127 \\
8    & 36& 133& 206& 238& 250& 254& 255& 255& 255& 255 \\
9    & 45& 221& 383& 464& 494& 506& 510& 511& 511& 511 \\
10   & 55& 364& 709& 894& 974& 1006& 1018& 1022& 1023& 1023 \\ \hline
\end{tabular}

\smallskip {\bf Table 1}
\end{center}

\medskip
\noindent From the definition of the $d$-substrings follows that
$$ N(k,d) = N(k,d+1), \qquad \hbox{for}\quad d\ge k-1 $$
 but
$$ N(k,k-1) = 2^k - 1 $$
and then
$$ N(k,d) = 2^k - 1, \qquad \hbox{for any}\quad   d \ge  k-1. $$
The following proposition gives the value of $N(k,d)$ in almost all cases:

\begin{proposition} {\em [3].} For $k \ge  2d-2$  we have
$$ N(k,k-d) = 2^k - (d-2)\cdot 2^{d-1} - 2.$$
\end{proposition}

The main step in the proof is based on the formula

$$N(k,k-d-1) = N(k,k-d) - d\cdot 2^{{\rm d-1}}.$$

The value of $N(k,d)$ can be also obtained by computing the number of
sequences of length $ k$ of $0's$ and $1's$, with no more than
$ d-1$ adjacent zeros. In such a sequence one 1 represents the presence, one
0 does the  absence  of a letter of the string in a given $d$-substring.
Let $b_{\rm k,d}$  denote  the  number  of
$k$-length sequences of zeros and ones, in which the first and last
position  is 1, and the number of adjacent zeros is at most $d-1.$
Then easily can be proved that

\medskip
 \hspace*{1 cm} $b_{k,d} = b_{k-1,d} + b_{k-2,d} + \ldots  +b_{k-d,d},$
                 for $ \;\;k > 1,$ \hfill \hspace*{1cm}$\; $

\hspace*{1 cm} $ b_{1,d} = 1,$

\hspace*{1 cm} $ b_{k,d} = 0$,  for all $k \le  0,$ \hfill \hspace*{1cm}$\;$

\medskip
\noindent because any such sequence of length $ k-i$  ($i=1,2,...,d$)
  can  be  continued  in
order to obtain a similar sequence of length $k$ in only one way
(by  adding  a sequence of the form $0^{i-1}1$ on the right).
For $b_{k,d}$ the following formula also
can be derived:
$$ b_{k,d} = 2b_{k-1,d} - b_{k-1-d,d}.$$
If we add one 1 or 0 in a internal position  (e.g  in  the $(k-2)^{th})$  of
each $ b_{k-1,d}$ sequences, then we obtain $2b_{k-1,d}$ sequences of length
$k$,  but  between
these $ b_{k-1-d,d}$ sequences will have $d$ adjacent zeros.

\medskip
The generating function corresponding to $b_{n,d}$ is
$$
B_{d}(z) = \sum_{n\ge 0}^{}{b_{n,d}z^n} =
 \frac{z}{1-z \cdots - z^d}= \frac{z(1-z)}{1-2z+z^{d+1}}.  $$
Adding zeros on the left and/or on the right to these sequences, we can obtain
the number $N(k,d)$, as the number of all these sequences. Thus
$$ N(k,d) = b_{k,d} + 2b_{k-1,d} + 3b_{k-2,d} + \cdots + kb_{1,d}. $$
($i$ zeros can be added in $i+1$ ways to these sequences: $0$ on the left and
$i$  on the right, $1$ on the left and $i-1$ on the right, and so on).

From the above formula, the generating function corresponding to the
complexities $N(k,d)$ can be  obtained as a product of the two generating
functions $B_d(z)$ and $A(z) = \sum_{n\ge 0}^{}{nz^n}= 1/(1-z)^2$, thus:
$$ N_d(z)=\sum_{n\ge 0}^{}{N(n,d)z^n} = \frac{z}{(1-z)(1-2z+z^{d+1})}.$$

\section{The 1-complexity}

We shall use the term {\em complexity} instead of the
{\em 1-complexity} and the  notation {\bf K}({\bf p}) instead of
${\bf K}_{1}({\bf p})$. A $k$-length string {\bf p}  over  an
$n$-letter  alphabet  has maximal complexity if

$${\bf K}({\bf p}) = \sum_{i=1}^{k}{ \min  ( n^i, k-i+1)}.$$
In the following we give some results which can be proved immediately
(in all cases ${\bf p} \in  X^k)$:

$a)\;\;\;k \le  {\bf K}({\bf p}) \le  \displaystyle\frac{k(k+1)}{2}$.

$b)\;\;\; {\rm For \;a \;trivial\; string }\;\; {\bf p} = a^k,\quad
{\bf K}({\bf p}) =  k.$

$c)\;\;\; {\rm If } \;\; x_{k} \neq  x_{i} \;{\rm for }\;
 i = 1,2,\cdots ,k-1,$ then
 $${\bf K}(x_1 x_2\cdots x_k) = k +{\bf K}(x_1 x_2\cdots x_{k-1})$$.

$d)\;\;\; {\rm If\; {\bf p} \;is \;not\; a\; trivial\; string,\; then}
\quad 2k-1 \le  {\bf K}({\bf p}) \le  \displaystyle\frac{k(k+1)}{2}. $

$ e) \;\;\;{\rm If} \;{\bf p} =  a^{i-1}ba^{k-i} \;{\rm for\; a \;fixed}
 \;i \;\;(1 \le  i \le  \lfloor k /2 \rfloor)$ then
  $${\bf K}({\bf p}) = (i+1) k - i^2.$$

$ f)\;\;\; {\rm If\; {\bf p}\; has\; at \;least \; \ell\; different\;
 letters \;then} \quad {\bf K}({\bf p}) \ge  k\ell
    -\displaystyle\frac{\ell (\ell -1)}{2}$.

(For the string $ a_1 a_ 2 \cdots a_{l-1}b^{k-l}$   with
$a_i\neq a_j$ for $i\neq j$, and $a_i\neq  b$ we  have
equality in the above formula).

$ g)\;\;\; {\rm If \;{\bf p}} \; \in X^k, {\bf q} \in Y^m \;{\rm and}
\; X \cap  Y = \emptyset $  then  
 $${\bf K}({\bf pq}) = {\bf K}({\bf p}) + {\bf K}({\bf q})  + km.$$

$ h)\;\;\; {\rm If}\; {\bf p}$ has only different letters then
    
  \hspace*{0.5 cm} {\bf K}({\bf p}) =$\displaystyle\frac{k(k+1)}{2}, $

  \hspace*{0.5 cm} ${\bf K}({\bf pp}^R) =2k^2,\;\;$ where {\bf p}$^R$  is
 the reverse string of {\bf p},

  \hspace*{0.5 cm} ${\bf K}({\bf p}^n) =
\displaystyle\frac{k(k+1)}{2}+(n-1)k^2,\;\;$
 where ${\bf p}^n$ is {\bf p} concatenated $n$ times.

$i)\;\;\; {\bf K}(x_1 x_2 \cdots x_k x_1 x_2\cdots x_n) =
\displaystyle\frac{k(k+1)}{2} + nk$ \quad for $1\le n\le k$,
$ x_i\neq x_j$ for $i\neq j.$

\bigskip
There arise the following two problems:
\par\medskip
1. {\em Find a minimal length string with a given complexity.}
\par\smallskip
\noindent This problem always has solution. (If the complexity is $C$, then
in the  worst case the string consisting of $C$ identical letters represents
a trivial solution).
\par\medskip
2. {\em Find a $k$-length string with a given complexity, if it exists.}
\par\smallskip
These problems can be solved by  a {\em branch-and-bound}-type  algorithm.  We
shall construct a tree in which each node is a string. The root is a letter of
the alphabet. Each node (i.e. each string) is obtained from  its  parent  node
by adding a new letter of the alphabet. The contruction will be continued at a
node if its complexity is less than the given complexity, or in  the  case  of
the second problem only if its length is also less than $k.$ This algorithm can
be improved by omitting some branches, which do not  produce  essentially  new
strings, e.g. if we have a four letter alphabet, then the strings $abd$ and
$abc$
are isomorphic (differ only some letters, but not the form). This can be given
by the following recursive algorithm. Let $a_1,a_2,\cdots ,a_n$ be the letters
of the alphabet, $k$ the length and $C$ the  desired  complexity.  The  symbol
 "+"  will denote the concatenation of a string with a letter, $|w|$ the length
  of  string $ w$. The algorithm starts with {\em generate("$a_1$")}.

\medskip
 {\em  generate (w):

\hspace*{0.5 cm} 
     if complexity (w) $< C$ and $| w|< k$

\hspace*{0.5 cm} 
    then   for $i := 1,2,\cdots ,k$ do generate( w + "a$_i$").

\hspace*{0.5 cm} 
    else   if complexity (w) = C and  $| w|  = k$  then write (w).}

\medskip
Of course, if $C < k$ or $ C > k(k+1)/2$ or doesn't exist a string
with the desired
complexity and length, then this algorithm  produces  nothing.  To  solve  the
first problem, we omit the restriction on length in the above algorithm.

If there is always a string with a given  complexity,  the  question  is:
there  exists  a  nontrivial  string  with  a  given  complexity  or  not?  (A
nontrivial string contains at least two different letters). The answer is yes,
except some cases.

\begin{proposition}If $C$  is a natural number different from 1, 2 and
 4,  then there exists a nontrivial string of complexity equal to $C.$
\end{proposition}

{\em Proof}. To prove this proposition we give the complexity of the  following
$k$-length strings:
\par\medskip
\hspace*{1cm}$ {\bf K}(a^{k-1}b) = 2k-1$ \hfill for $ k \ge 1\hspace*{1cm}\;$

\hspace*{1cm}$ {\bf K}(ab^{k-3}aa)  = 4k-8$ \hfill for $k \ge 4\hspace*{1cm}\;$

\hspace*{1 cm}$ {\bf K}(abcd^{k-3}) = 4k-6$ \hfill for $k \ge 3\hspace*{1cm}\;$

These can be proved immediately from the definition of
the complexity.

 1. If $C$ is odd then we can write $ C = 2k-1$ for a given $k.$
 From this $k =(C+1)/2$
results, and the string $a^{k-1}b$, has complexity $C.$

2. If $C$ is even, then $C = 2\ell $.

\hspace*{0.5 cm} 2.1. If $\ell  = 2h$, then $4k-8 =C$ gives $4k-8 = 4h$,
 and from this $k=h+2$ results.
The string $ ab^{k-3}aa$  has complexity $C.$

\hspace*{0.5 cm} 2.2. If $\ell  = 2h+1$ then $4k-6 = C$ gives
$4k-6 = 4h+2,$ and from this $k=h+2$ results.
The string $abcd^{k-3}$  has complexity $C.\;$  \blacksquare

\medskip
In the proof we have used more than two letters in a string only  in  the
case of the numbers of the form $4h+2$ (case 2.2 above).
The new question is, if
there exist always nontrivial strings formed only of two letters with a  given
complexity. The answer is yes anew. We must prove this only for the numbers of
the form $4h+2.$ If $C = 4h+2$ and $C \ge  34,$ we use the followings:

\medskip
\hspace*{1 cm}${\bf K}(ab^{k-7}abbabb)= 8k-46,$   
             \hfill for $k \ge 10,\hspace*{1cm}$

\hspace*{1 cm}${\bf K}(ab^{k-7}ababba) = 8k-42,$ 
              \hfill for $k \ge 10.\hspace*{1cm}$

\medskip
If $h = 2s$, then $8k-46 = 4h+2$ gives $k = s+6,$ and the string
$ab^{k-7}abbabb$ has complexity $4h+2.$

If $h = 2s+1,$ then $8k-42 = 4h+2$ gives $k = s+6,$ and the string
$ab^{k-7}ababba$  has complexity $4h+2.$ For $C < 34$ only 14, 26 and 30 are
feasible. The  string $ab^4a$
has complexity 14, $ab^6a$ complexity 26, and $ab^5aba$ complexity 30. Easily
can be proved, using a tree like in the above algorithm, that for 6, 10, 18
and  22 such strings does not exist. Then the following is true.

\begin{proposition}
If $C$ is a natural number different from 1, 2, 4, 6,
 10, 18 and 22, then there exists a nontrivial string formed only of two 
letters, with the given complexity $C.$
\end{proposition}

In relation with the second problem a new one arises: How many strings of
length $k$ and complexity $C$ there exist? For small $k$ this problem can be
studied exhaustively. Let $X$ be of $k$ letters, and let us consider all
strings of length $k$ over $X.$ By a computer program we have got Table 2,
which  contains the frequency of strings with a given length and complexity.

\medskip
\hrule

\begin{tabular}{lrrclrrrr}
{\em length=2}   &   &   & \hspace*{0.5 cm} & {\em length=3}  &   &   &   &   \\
{\em complexity} & 2 & 3 &                  & {\em complexity} & 3 & 4 & 5 & 6  \\
{\em frequency}  & 2 & 2 &                  & {\em frequency} &3 & 0 & 18 & 6
\end{tabular}

\smallskip
\begin{tabular}{lrrrrrrr}
{\em length=4}   &   &   &   &   &    &   &     \\
{\em complexity} & 4 & 5 & 6 & 7 &  8 & 9 & 10   \\
{\em frequency } & 4 & 0 & 0 & 36 &48 & 144 & 24   \\
\end{tabular}

\smallskip
\begin{tabular}{lrrrrrrrrrrr}
{\em length=5 }  &   &   &   &    &   &    &    &    &    &    & \\
{\em complexity} & 5 & 6 & 7 &  8 & 9 & 10 & 11 & 12 & 13 & 14 & 15  \\
{\em frequency } & 5 & 0 & 0 &  0 & 60& 0  & 200& 400&1140&1200&120   \\
\end{tabular}

\smallskip
\begin{tabular}{lrrrrrrrrr}
{\em length=6}   &   &   &   &   &    &    &    &    &           \\
{\em complexity} & 6 & 7 &  8& 9 & 10 & 11 & 12 & 13    \\
{\em frequency}  & 6 & 0 & 0 & 0 & 0  & 90 & 0  & 0   \\  \\
         & 14  & 15 &16 &17  &18   &19   &  20 &  21 &         \\
         & 300  &990 &270&5400&8280 &19800&10800& 720 &        \\
\end{tabular}

\hrule

\smallskip
\centerline{\bf Table 2.}

\medskip
Let $|X|=k$ and let $f_k(C)$ denote the frequency of the $k$-length
strings over $X$ having a complexity $ C$. Then the following proposition is
true.

\begin{proposition}

\hspace*{1cm} $f_k(C) = 0$\qquad
            if $C<k$ or $ C>\displaystyle\frac{k(k+1)}{2}, $\\[2pt]

\hspace*{1 cm} $f_k(k) = k$,\\[2pt]

\hspace*{1 cm} $f_k(2k-1) = 3k(k-1),$ \\[2pt]

\hspace*{1 cm} $f_k\left( \displaystyle\frac{k(k+1)}{2} -1\right) =
  \displaystyle\frac{k(k-1)k!}{2}, $\\[2pt]

\hspace*{1 cm} $f_k\left( \displaystyle\frac{k(k+1)}{2} \right) =
  k! $
\end{proposition}

{\em Proof.} The first two and the last ones are  evident.  Let  us  prove  the
third. If the complexity of a $k$-length string is $2k-1,$ then
it must contain exactly two substrings of length $1,2,\cdots ,k-1$, and only one
of the length $k$, and must be formed of two letters. (If it contains 3 letters
than the complexity is $\ge 3k-3$, see the property $f).$ ) In this case  the
2-lentgh substrings can be only $aa, ab$ or $aa, ba$ or $ab, ba$, and with
these only strings of the form $a^{k-1}b$, $ba^{k-1}$ and $(ab)^{k/2}$ (if $k$
is even) or $(ab)^{(k-1)/2}a$ (if $k$ is odd) can be generated. In every case
the two letters can be chosen in $k(k-1)$ ways, and because of the three above
possibility $f_k(2k-1)=3k(k-1)$.

The last but one comes from the following: $k$ letters can form $k!$ different
$k$-length strings of maximal complexity, and the complexity of such a string
can be diminished by one  if we replace a letter by another already being
present in that string. We can choose a position for one already  given in
$k(k-1)$ ways, and because of the symmetry of the letters in these positions,
the number of new strings is $k!k(k-1)/2$. \blacksquare

As regards the distribution of the frequaency $0$, we can prove the following.

\begin{proposition}

\centerline{If $C=k+1, k+2, \cdots , 2k-2,\;\;$ then $\; f_k(C)=0.$}
\centerline{If $\;C\;=\;2k, 2k+1, \cdots , 3k-5,\;\;\;$ then $\; f_k(C)=0.\;$}
\end{proposition}

{\em Proof. } The complexity of the trivial $k$-length string is $k$, and this contains
only one letter $k$ times. If in such a string we replace one or more
letters by a new one, the number of substrings of any length, except the whole
string, will increase by at least one. Then the complexity will be at least
$2k-1$, and there are no strings with complexity between $k$ and $2k-1$. To
prove the second formula, we use the following, easy to see assertion:
{\em if a k-length string has n i-length  substrings, then it has at least
{\rm min}(n,k-i+1) $\;$ (i+1)-length substrings.}

By replacing a letter with a new one in the strings of complexity $2k-1$, we
obtain at least complexity $3k-3$. If we replace one $a$ (or more) with one $b$
(or more), or inversely, but not to obtain a trivial string, and keeping
the length, the number of 2-length substrings will increase by 3, and
by the above assertion will increase the number of $3-, 4-, \cdots ,
(k-2)-$length substrings. Then the complexity will be at least   $2+3(k-3)+2+1$
which is $3k-4. \blacksquare$

Strings of length $k$ may have complexity between $k$ and $k(k+1)/2$. Let us
denote by $b_k$ the least number for which

\medskip
\centerline{$f_k(C) \ne 0\;\;$ for all $C$ with $\;\;b_k \le C \le
\displaystyle\frac{k(k+1)}{2}$. }

\medskip
The number $b_k$ exists for any $k$ (in the worst case it may be equal to
$k(k+1)/2$).
In the Table 2 we can see that $b_3=5$, $b_4=7$, $b_5=11$ and $b_6=14.$

\smallskip
We give the following conjecture:

{\sc Conjecture.}
If $k= \displaystyle\frac{\ell (\ell +1)}{2}+2+i$,
where $\ell \ge 2$ and $0\le i\le \ell $ then
$$ b_k = \frac{\ell (\ell^2 -1)}{2}+3\ell+2+i(\ell+1).\;\; \blacksquare   $$

We can easily see that $f_k(b_k) \ne 0$ for $k\ge 5$, because of
${\bf K}(ab^{k-\ell}ab^{\ell -2})$ $=b_k$.

\section*{Conclusions}

We have studied the $d$-complexity of strings,  which  is  defined  as  the
number of all distinct $d$-substrings of it. The concept of the $d$-substring
is a generalization of that of the substring: not  only a contiguous part of a
string can be chosen as substring, but parts which have distance between them
no greater than $d$. The $d$-complexity of strings with different letters only,
can be computed by a Fibonacci-type sequence. Proposition 1 gives a formula
for this complexity in almost all cases.

The 1-complexity is studied in detail. In propositions 2 and 3 we prove
that, except some cases, a string with a given complexity can be associated to
any natural number. The frequency of strings with a given complexity is also
considered. It is conjectured that if we consider strings of length $k$, there
exists a value between $k$ and $k(k+1)/2$ from which 0 frequency no more
exists.

\end{document}